\documentclass[11pt]{article}
\usepackage{amsmath,amssymb,amsfonts}
\usepackage[english]{babel}
\makeatletter
\@addtoreset{equation}{section}
\makeatother

\def\nn{\nonumber} \def\bd{\begin{document}} \def\ed{\end{document}}
\def\ds{\documentstyle}
\let\bm=\bibitem
\newcommand{\be}{\begin{equation}}
\newcommand{\ee}{\end{equation}}
\newcommand{\bea}{\setlength\arraycolsep{2pt} \begin{eqnarray}}
\newcommand{\eea}{\end{eqnarray}}
\newcommand{\hoch}[1]{$\, ^{#1}$}
\def\p{\partial}
\title{\large {\bf A Komar-like integral for mass and angular momentum
of asymptotically AdS black holes in Einstein gravity}}
\date{}

\author{Jun-Jin Peng$^{1,2}$\footnote{corresponding author: pengjjph@163.com},
\quad Chang-Li Zou$^{1}$,
\quad Hui-Fa Liu$^{1}$\\ \\
\small \sl $^1$School of Physics and Electronic Science,\\
\small \sl Guizhou Normal University,\\
\small Guiyang, Guizhou 550001, People's Republic of China; \\
\small \sl  $^{2}$Guizhou Provincial Key Laboratory of Radio Astronomy and
Data Processing, \\
\small \sl Guizhou Normal University, \\
\small Guiyang, Guizhou 550001, People's Republic of China
}

\begin{document}

\maketitle
\vspace{5pt}

\begin{center}
\textbf{Abstract}
\end{center}

The purpose of this paper is to enhance the conventional Komar integral
to asymptotically anti-de Sitter (AdS) black holes. In order to do so, we
first obtain a potential that is the linear combination of the usual Komar
potential with two third-order derivative terms generated by the action of
the d'Alembertian operator and the exterior derivative upon a Killing vector.
Then this higher-order corrected potential is extended to the Einstein
gravity with a negative cosmological constant, yielding the potential that
is the linear combination of the usual Komar one with it acted on by
the d'Alembertian. The surface integral of the improved Komar
potential can serve as a formula for conserved charges of asymptotically
AdS spacetimes. Finally, we make use of such a formula to compute
the mass and the angular momentum of Schwarzschild-AdS black holes,
regular AdS black holes, asymptotically AdS Kerr-Sen
black holes, Kerr-NUT-AdS black holes, and Kerr-AdS black holes in
arbitrary dimensions. The results coincide with the ones in the literature.

%\textbf{Keywords}:

%\textbf{PACS}: 04.50.Kd; 04.50.-h

%%%%%%%%%%%%%%%%%%%%%%%%%%%%%%%%%%%%%%%%%%%%%%%%%%%%%%%%%%%%%%%%%%%%%%%%%
\voffset=-.90pt
\vspace{40pt}

%%%%%%%%%%%%%%%%%%%%%%%%%%%%%%%%%%
\section{Introduction}\label{one}
%%%%%%%%%%%%%%%%%%%%%%%%%%%%%%%%%%

There has been a great deal of interest in the asymptotically anti-de Sitter
(AdS) black holes over the past several decades, particularly since the
discovery of the well-known anti-de Sitter/conformal field theory (AdS/CFT)
correspondence \cite{MAACFT}. Up to now, rotating black hole solutions
with a cosmological constant have been found in different dimensions within the
framework of the Einstein gravity theory. All of them satisfy exactly the
vacuum Einstein field equation in the presence of the cosmological constant.
In 1968, Carter first presented a
generalization of the four-dimensional rotating Kerr black hole by including
the cosmological constant \cite{4DKeAdSsolu}. Since this black hole possesses
asymptotically de Sitter (dS) or AdS boundary conditions, it is usually
referred to as Kerr-dS or Kerr-AdS black hole in the literature. After this,
by introducing a NUT charge parameter and acceleration in the Kerr-AdS metric,
the four-dimensional accelerating Kerr-NUT-AdS black hole solution was found
in \cite{4DKNUTC,4DKNuPK}. Many years later, Hawking, Hunter and
Taylor-Robinson found the five-dimensional generalization of the
four-dimensional Kerr-(A)dS black hole, as well as the solutions with just
one nonzero angular momentum parameter in all dimensions \cite{5DsoluHHT}.
Subsequently, Gibbons, L\"{u}, Page and Pope further constructed the general
Kerr-(A)dS black holes with arbitrary angular momenta in all higher
dimensions \cite{GLuPP1,GLuPP2}, which can be regarded as the extensions of
the Ricci-flat rotating Myers-Perry black holes \cite{MyersP} in the
appearance of the cosmological constant. Soon afterwards the
higher-dimensional NUT charge generalizations for the Kerr-(A)dS black
holes were obtained in \cite{CLPKNuT}.

For the sake of understanding the asymptotically AdS black holes or the
spacetimes with other asymptotical structures, in particular, their first
law of thermodynamics together with other thermodynamic properties of
relevance to energy and angular momentum, a question desired to address is
to seek proper definitions for the conserved charges. Till now, some
research has been devoted to this question from different perspectives, and
a lot of progress has already been made, enriching our understanding on the
conserved charges of gravity theories. For the asymptotically AdS spacetimes,
there have existed a number of methods that can be adopted to define their
conserved charges in
the literature, for example, the well-known covariant phase space approach
\cite{LeeWald,IyWald,WalZo}, the KBL superpotential method
\cite{KBLSu1,KBLSup}, the conformal definition of conserved charges proposed
by Ashtekar, Magnon and Das (AMD) \cite{AMDmass,AMDmass2}, the Brown-York
approach \cite{BrYor}, the counterterm
method \cite{CoutBK,CoutKLS}, the Abbott-Deser-Tekin (ADT) formalism
\cite{AbbottD1,AbbottD2,DeserT1,DeserT2} together with its off-shell
generalization \cite{KimKY}, the topological regularization method
\cite{ToPReGR,ACOTZ,Olen4Dker}, the Barnich-Brandt-Compere (BBC) formalism
\cite{BarnB,Barn2,BarnichC,BCintegC},
the field-theoretic approach \cite{PetrFT1,PetLogr,PetPre}, and
the method proposed in \cite{ObukRub,ObPPR}.

By contrast with the aforementioned methods, the so-called Komar integral,
which is a surface integral with respect to a 2-form potential made up of the
first-order derivative of a Killing vector field \cite{Komar}, may be thought
of as the simplest formalism for conserved charges. Thanks to such a merit,
it is of great convenience to use the Komar integral
to evaluate the mass and the angular momentum of various solutions of gravity
theories, particularly for the ones with asymptotically flat structure.
Naturally, one expects that the simple formulation is also applicable to
asymptotically AdS spacetimes. Nevertheless, when the Komar integral is
applied to compute the mass of the asymptotically AdS black holes  (for
instance, see the calculations of the mass for the static and spherically
symmetric Schwarzschild-AdS black holes in Subsection \ref{three1}),
unfortunately, it breaks down, arising from that the result turns out to be
infinite. This implies that there must exist divergent terms at infinity.
In this sense, the regularization of such terms requires at least that the
usual Komar integral has to be modified to keep on its success in the
asymptotically AdS spacetimes. What is more, although the original Komar
integral can produce the physically meaningful mass together with the angular
momentum of the asymptotically flat rotating black holes, such as the
Myers-Perry black holes in all dimensions, the expression with respect to
the mass differs from that for the angular momentum by a factor two in form
\cite{Katfac2}. Therefore, to obtain a unified formulation for both the
energy and the angular momentum, the original Komar integral should be
improved.

In the present paper, we pursue the goal of providing a natural generalization
of the conventional Komar integral so that it can work for the conserved charges
of various spacetimes with an AdS asymptotic in the framework of the Einstein
gravity theory. In order to do so, a proper modification to the ordinary
Komar potential \cite{Komar} has to be made. Inspired by the expression of
the Komar potential in differential forms, being the exterior derivative of
a 1-form Killing vector field, a good candidate for the generalization of
the Komar potential may be the one encompassing the higher-order derivatives
of the Killing vector.
Fortunately, we find that the approach to generate conserved currents associated
with an arbitrary vector field put forward in the works \cite{PenCTP,PenZ}
can assist us to accomplish the higher-order corrections to the Komar potential.
The relevant idea behind these works goes as follows. If both the current and the
vector field are treated as 1-forms, by letting the degree-preserving
d'Alembertian operation together with the codifferential and the exterior
derivative (both of them have to be in pair) act on the 1-form vector field,
one is able to obtain a series of 1-forms being the even-order derivatives
of the vector field. By means of the linear combination of all the 1-forms,
then conserved currents associated with the vector can be constructed.

Within what follows, on basis of the strategy of generating conserved
currents in \cite{PenCTP,PenZ}, we obtain an identically conserved current
consisting of the second- and fourth-order derivatives of an arbitrary
Killing vector
field. In terms of this current, a potential is immediately read off, which
is the linear combination of the usual Komar potential with two third-order
derivative terms generated by the action of the d'Alembertian and the exterior
derivative on the 1-form Killing vector. Due to the structure of the potential, it
can be regarded as the generalization of the usual Komar potential with
higher-order corrections. Accordingly, we refer to the surface integral of this
potential as a Komar-like integral. Furthermore, the generalized Komar potential is
extended to the Einstein gravity theory endowed with the
Einstein-Hilbert-$\Lambda$ Lagrangian, yielding a simplified and specific
potential. The surface integral with respect to the Komar-like potential
gives rise to a unified definition for the energy and the angular momentum
of asymptotically AdS black holes. As applications, we compute the mass and
the angular momentum of the Schwarzschild-AdS black holes in arbitrary
dimensions, the four-dimensional regular AdS black holes \cite{FWreAdS},
the four-dimensional asymptotically AdS Kerr-Sen black holes \cite{KSAdS4},
the Kerr-NUT-AdS black holes \cite{4DKNUTC,4DKNuPK,CLPKNuT}, and the Kerr-AdS
black holes in arbitrary dimensions \cite{GLuPP1,GLuPP2}.

The plan of the present paper is as follows. In section \ref{two}, we make
use of the d'Alembertian operator, the exterior derivative and the
co-exterior derivative to act on an arbitrary Killing vector field to produce
a conserved current with the linear combination of no more than fourth-order
derivative terms of this vector. Subsequently, by specifying the current to
Einstein gravity in the presence of a negative cosmological constant, we
obtain a
generalized Komar potential by means of the equation of motion and the
properties of Killing vectors. The surface integral of the potential
yields a formula of conserved charges for asymptotically AdS spacetimes.
Section \ref{three} is devoted to the applications of the formula for
conserved charges into the calculations for the mass and the angular
momentum of black holes with an AdS asymptotic. The last section is our
conclusions. In Appendix \ref{appendA}, to see more clearly the
contributions from the higher-order corrections to potentials, the
inclusions of more than third-order derivatives of the Killing vector
in the potentials are investigated.

%%%%%%%%%%%%%%%%%%%%%%%%%%%%%%%%%%%%%%%%%%%%%%%%%%%%%%%%%%%%%%%%%%%%%%%%
\section{General formalism}\label{two}
%%%%%%%%%%%%%%%%%%%%%%%%%%%%%%%%%%%%%%%%%%%%%%%%%%%%%%%%%%%%%%%%%%%%%%%%

In the present section, by following the work \cite{PenZ}, which provides a
way of constructing conserved currents starting from the action of
differential operators on an arbitrary vector field, we manage to find out
a generalized Komar potential for solutions with asymptotically AdS structure
in the context of $D$-dimensional Einstein gravity. Then the surface integral
with respect to the generalized potential will be adopted to define the
conserved charges such as the mass and the angular momentum. According to
the work \cite{PenZ}, the differential operators ${\rm d}$ and $\hat{\delta}$,
together with $\Box=\nabla^\mu\nabla_\mu$, participate in the construction of
the conserved current corresponding to the 1-form vector field. For convenience,
here we present their definitions through the action upon an
arbitrary $p$-form $F=(p!)^{-1}F_{\mu_1\cdot\cdot\cdot\mu_p}
{\rm d}x^{\mu_1}\wedge\cdot\cdot\cdot\wedge{\rm d}x^{\mu_p}$, under the metric
signature $(-,+,\cdot\cdot\cdot,+)$. The exterior
derivative ${\rm d}$ and the degree-preserving d'Alembertian $\Box$ are
defined in components through
$({\rm d}F)_{\mu_0\cdot\cdot\cdot\mu_p}
=(p+1)\nabla_{[\mu_0}F_{\mu_1\cdot\cdot\cdot\mu_p]}$
and $(\Box F)_{\mu_1\cdot\cdot\cdot\mu_p}
=\nabla^\rho\nabla_\rho F_{\mu_1\cdot\cdot\cdot\mu_p}$, respectively,
while the co-exterior derivative $\hat{\delta}$ is defined in terms of
the combination of the Hodge dual $\star$ and the exterior derivative,
taking the form $\hat{\delta}=(-1)^{pD+D+1}\star{\rm d}\star$. Here and
below the Hodge dual acts on the $p$-form $F$ as
$(\star F)_{\mu_1\cdot\cdot\cdot\mu_{D-p}}=
(p!)^{-1}F^{\nu_1\cdot\cdot\cdot\nu_p}
\epsilon_{\nu_1\cdot\cdot\cdot\nu_p\mu_1\cdot\cdot\cdot\mu_{D-p}}$
with the completely skew-symmetric Levi-Civita tensor given by
$\epsilon_{\mu_0\cdot\cdot\cdot\mu_{D-1}}=\sqrt{-g}
D!\delta^0_{[\mu_0}\cdot\cdot\cdot\delta^{D-1}_{\mu_{D-1}]}$. According
to the definitions, one directly obtains the identities
${\rm d}^2=0=\hat{\delta}^2$.
More details on ${\rm d}$, $\hat{\delta}$ and $\Box$ can be found in
\cite{PenCTP} and references therein. Furthermore, on basis of the
relationship between the generalized Komar potential and the potential
given in \cite{KasKI,KaRTinJ}, we shall propose a suitable substitute for
the Killing potential.

We start with the vacuum Einstein gravity theory in $D$ dimensions, equipped
with the Einstein-Hilbert-$\Lambda$ Lagrangian of the form
\be
\mathcal{L}_{EH}=\sqrt{-g}(R-2\Lambda)
\, . \label{EHLLag}
\ee
The variation of the Lagrangian (\ref{EHLLag}) with respect to the metric
tensor $g^{\mu\nu}$ gives rise to the gravitational field equation
\be
R_{\mu\nu}=-(D-1)\ell^2g_{\mu\nu}
\, , \label{Eqmotion}
\ee
where $\ell^{-1}$ stands for the radius of curvature for the maximally
symmetric AdS spaces (for the metric of the $D$-dimensional AdS space see
Eq. (\ref{SchAdS}) without the $m$ parameter in Subsection \ref{three1}),
related to the cosmological constant $\Lambda$ in the manner
\be
\ell^2=-\frac{2\Lambda}{(D-1)(D-2)}
\, . \label{elldef}
\ee
%\ell in IJMPA 35, 2050102 (2020)--->-\ell^2
In what follows, along the lines of the works \cite{PenCTP,PenZ},
we focus on giving a generalized Komar integral for the asymptotically AdS
solutions of the field equation (\ref{Eqmotion}), with the Riemann
curvature tensor at spatial infinity
\be
R^{\mu\nu}_{\rho\sigma}\big|_\infty=-2\ell^2
\delta^{[\mu}_{[\rho}\delta^{\nu]}_{\sigma]}
\, . \label{RiemSI}
\ee

Within the work \cite{PenZ}, it has been demonstrated that a conserved
current with respect to a vector field can be constructed through the operation
of the $(\Box,\hat{\delta},{\rm d})$ operators, as well as their combinations,
on such a vector if both the current and vector field are treated as 1-forms.
This is attributed to the fact that the second-order derivative operations
$\Box$, $\hat{\delta}{\rm d}$ and ${\rm d}\hat{\delta}$ are the three basic
differential operators leaving the form degree of any differential form
unchanged so that their action on the 1-form vector field yields a 1-form
as well. Inspired with this strategy of constructing currents, for
concreteness, we consider the conserved current
$\mathcal{J}^{(4th)}_{gen}(\xi)$ associated with an arbitrary Killing vector
$\xi$, generated only by acting the $(\Box,\hat{\delta},{\rm d})$ operators
together with their combinations upon $\xi$. Since
$\mathcal{J}^{(4th)}_{gen}(\xi)$ is expected to be a natural and simple
generalization of the usual Komar current
$J_{Komar}=-\hat{\delta}{\rm d}\xi$ \cite{Komar}, it is reasonable
to assume that $\mathcal{J}^{(4th)}_{gen}(\xi)$ is still linear in $\xi$.
Apart from this, when the Komar integral based on the current $J_{Komar}$
is applied to compute the mass of asymptotically AdS black holes, it has
been shown before that such an integral fails to yield a finite physical
result because of the appearance of
the divergent terms at spatial infinity, while the higher-order derivatives
of the Killing vector have great potential to cancel out such divergent terms.
Thus it is supposed that higher-order derivative terms of $\xi$ are contained
within $\mathcal{J}^{(4th)}_{gen}(\xi)$. However, the simplest and most
natural higher-order derivative generalization to $J_{Komar}$ is to merely bring
fourth-order derivatives of $\xi$ into $\mathcal{J}^{(4th)}_{gen}(\xi)$ (for discussions
on this see Appendix \ref{appendA}). Consequently, under the aforementioned
assumptions, the current
$\mathcal{J}^{(4th)}_{gen}(\xi)$, consisting of terms proportional to no more than fourth-order
derivatives of $\xi$, has the most general structure
\be
\mathcal{J}^{(4th)}_{gen}(\xi)
=J^{(2th)}_{gen}(\xi)+J^{(4th)}_{gen}(\xi)
\, , \label{CalJ4gen}
\ee
where the second-order derivative 1-form $J^{(2th)}_{gen}(\xi)$, which was
utilized to reformulate the conventional Komar current in \cite{PenZ}, is
given by
\be
J^{(2th)}_{gen}(\xi)
=\lambda_{11}\Box\xi
+\lambda_{12}\hat{\delta}{\rm d}\xi
+\lambda_{13}{\rm d}\hat{\delta}\xi
\, , \label{J2gen}
\ee
while the 1-form $J^{(4th)}_{gen}(\xi)$ consisting of the fourth-order
derivatives of the Killing vector reads as follows:
\bea
J^{(4th)}_{gen}(\xi)
&=&\lambda_{21}\Box\hat{\delta}{\rm d}\xi
+\lambda_{22}\hat{\delta}{\rm d}\hat{\delta}{\rm d}\xi
+\lambda_{23}{\rm d}\hat{\delta}{\rm d}\hat{\delta}\xi\nn \\
&&+\lambda_{24}\Box\Box\xi
+\lambda_{25}\Box {\rm d}\hat{\delta}\xi
+\lambda_{26}{\rm d}\Box \hat{\delta}\xi  \nn \\
&&+\lambda_{27}{\rm d}\hat{\delta}\Box \xi
+\lambda_{28}\hat{\delta}{\rm d}\Box \xi
+\lambda_{29}\hat{\delta}\Box {\rm d}\xi
\, . \label{J4thgen}
\eea
In Eqs. (\ref{J2gen}) and (\ref{J4thgen}),
$\lambda_{1i}$'s and $\lambda_{2j}$'s are constant parameters, and the
coderivative $\hat{\delta}$ has to be paired with the exterior derivative
${\rm d}$ to ensure that the action of the operators preserves the form degree
of the 1-form Killing vector field. Although $\mathcal{J}^{(4th)}_{gen}(\xi)$ appears
in a complicated form, by substituting the identities $\hat{\delta}\xi=0$ and
$\hat{\delta}{\rm d}\xi=2\Box\xi$ for Killing vectors into the above equation,
we simplify it as
\bea
\mathcal{J}^{(4th)}_{gen}(\xi)
&=&\left(\frac{\lambda_{11}}{2}
+\lambda_{12}\right)\hat{\delta}{\rm d}\xi
+(2\lambda_{22}+\lambda_{28})
\hat{\delta}{\rm d}\Box\xi\nn \\
&&  +\lambda_{29}\hat{\delta}\Box {\rm d}\xi
+\big(2\lambda_{21}+\lambda_{24}\big)\Box\Box\xi
\, . \label{CalJ4gen2}
\eea
To guarantee that $\hat{\delta}\mathcal{J}^{(4th)}_{gen}(\xi)
=\big(2\lambda_{21}+\lambda_{24}\big)\hat{\delta}\Box^2\xi
=0$ holds for an arbitrary Killing vector $\xi$, it is demanded that
$\lambda_{24}=-2\lambda_{21}$. In such a setting, for convenience, the
identically conserved current $\mathcal{J}^{(4th)}_{gen}(\xi)$ is recast into the
following form
\be
\mathcal{J}^{(4th)}(\xi)=
k_1\hat{\delta}{\rm d}\xi
+2k_2\hat{\delta}{\rm d}\Box\xi
+k_3\hat{\delta}\Box {\rm d}\xi
\, , \label{CalJ4th}
\ee
with arbitrary constant parameters $k_i$'s. As usual, in terms of the
relation between the current and potential
$\mathcal{J}^{(4th)}(\xi)=-\hat{\delta}\mathcal{K}^{(3th)}(\xi)$,
the 2-form potential $\mathcal{K}^{(3th)}(\xi)$ corresponding to the
$\mathcal{J}^{(4th)}(\xi)$ current is straightforwardly read off as
\be
\mathcal{K}^{(3th)}(\xi)=
-k_1{\rm d}\xi
-2k_2{\rm d}\Box\xi
-k_3\Box {\rm d}\xi
\, . \label{CalK3th}
\ee
In contrast with the ordinary Komar potential $K_{Komar}={\rm d}\xi$ being
the first-order derivative of the Killing vector \cite{Komar}, here the
2-form $\mathcal{K}^{(3th)}(\xi)$ differs from $K_{Komar}$ by encompassing
two additional third-order derivative terms ${\rm d}\Box\xi$ and
$\Box{\rm d}\xi$. Such terms are generated through the action of the
differential operators on the 1-form Killing vector field, which is in
accordance with the spirit of constructing the original Komar potential.
Naturally, $\mathcal{K}^{(3th)}(\xi)$ can be regarded as the higher-order
derivative generalization of the ordinary Komar potential. Furthermore,
by making use of the Weitzenb\"{o}ck identity \cite{PenCTP}, we arrive at
\bea
\mathcal{K}^{(3th)}(\xi)
&=&-k_1{\rm d}\xi-2(k_2+k_3){\rm d}\Box\xi \nn \\
&&-2k_3R_{\mu}^{\rho}
\nabla_{\rho}\xi_{\nu}{\rm d}x^\mu\wedge{\rm d}x^\nu\nn\\
&&+k_3R_{\mu\nu}^{\rho\sigma}
\nabla_{\rho}\xi_{\sigma}{\rm d}x^\mu\wedge{\rm d}x^\nu
\, . \label{CalK3th2}
\eea
Here it should be emphasized that $\mathcal{K}^{(3th)}(\xi)$ in
Eq. (\ref{CalK3th2}) is general and it is irrelevant to equations of motion
for fields. The constant parameters $k_1$, $k_2$ and $k_3$ can be determined
by specifying $\mathcal{K}^{(3th)}(\xi)$ to a gravity theory under
consideration. All of them at least guarantee that the conserved charges
defined in terms of the 2-form $\mathcal{K}^{(3th)}(\xi)$, such as the
mass and the angular momentum, are convergent at infinity. This will be
demonstrated in the following part of this section, as well as within
Appendix \ref{appendA}.

With the generic Komar-like potential $\mathcal{K}^{(3th)}(\xi)$ in hand,
we concentrate on figuring out the potential with respect to the Einstein
gravity characterized by Lagrangian (\ref{EHLLag}). In the framework of this
gravity theory, under the equation of motion (\ref{Eqmotion}), as well
as the identity $\Box\xi=(D-1)\ell^2\xi$ arising from the relationship
$\nabla_\rho\nabla_\mu\xi_\nu=-R_{\mu\nu\rho\sigma}\xi^\sigma$ and the
field equation (\ref{Eqmotion}), the potential $\mathcal{K}^{(3th)}(\xi)$
turns into
\be
\mathcal{K}^{(3th)}_{AdS}=c_2\left({\rm d}\xi
+c_1R_{\mu\nu}^{\rho\sigma}
\nabla_{\rho}\xi_{\sigma}{\rm d}x^\mu\wedge{\rm d}x^\nu \right)
\, , \label{CalK3gr}
\ee
with $c_2=-\big[k_1+2(D-1)k_2\ell^2\big]$ and $c_1=k_3/c_2$. At this stage
there are actually only two undetermined constant parameters. As what will
be demonstrated in Appendix \ref{appendA}, the $c_1$ parameter can be
completely fixed by analyzing the asymptotical behaviour of the spacetimes.
For asymptotically AdS spacetime metrics obeying the falloff condition
(\ref{RiemSI}), due to the fact that the determinant of the metric suffers
from divergence at spatial infinity, the necessary but not the sufficient
condition for the potential $\mathcal{K}^{(3th)}_{AdS}$ to satisfy is that
it has to vanish at spatial infinity to
guarantee that the conserved charges defined in terms of the surface
integrals of $\mathcal{K}^{(3th)}_{AdS}$ are finite. As a consequence,
with the help of the equation
$\mathcal{K}^{(3th)}_{AdS}\big|_\infty=c_2\big(1-2c_1\ell^2\big){\rm d}\xi$,
such a condition gives rise to
\be
c_1=\frac{1}{2\ell^2}
\, . \label{kval1}
\ee
Under the $c_1$ parameter given by Eq. (\ref{kval1}), it will be shown in the
next section that the divergent Komar integral for the asymptotically AdS
black holes can be indeed regularized by the higher-order derivative
term $\Box {\rm d}\xi$.
Meanwhile, on the other hand, it seems not to be crucial for the determination
of the global factor $c_2$ attributed to that it can be incorporated
into the formula for the conserved charges. Actually, when the surface
integral in terms of $\mathcal{K}^{(3th)}_{AdS}$ is adopted to
calculate the conserved charges of black holes, for instance, the mass of
Schwarzschild-AdS black holes, in order to cover smoothly the results in the
absence of the cosmological constant or to reproduce the same mass as
that via other standard methods for conserved quantities, such as
the covariant phase space approach \cite{LeeWald,IyWald,WalZo}, the AMD
formalism \cite{AMDmass,AMDmass2}, the ADT method
\cite{AbbottD1,AbbottD2,DeserT1,DeserT2} and the BBC formalism
\cite{BarnB,Barn2,BarnichC,BCintegC}, for example, it will be confirmed
in the following subsection (\ref{three1}) that the $c_2$
parameter is presented by
\be
c_2=\frac{1}{2(D-3)}
\, . \label{kval2}
\ee
Consequently, by substituting Eqs. (\ref{kval1}) and (\ref{kval2}) into
Eq. (\ref{CalK3th}), we obtain the generalized Komar potential
$\mathcal{K}$, given by
\be
\mathcal{K}
=\frac{1}{2(D-3)}\left({\rm d}\xi
-\frac{\Box {\rm d}\xi}{2\ell^2}\right)
\, , \label{CalKAdS}
\ee
which is the linear combination of the ordinary Komar potential
$K_{Komar}$ with its second-order derivative yielded under the action of
the degree-preserving d'Alembertian operation. Here the Komar-like potential
$\mathcal{K}$ is just the natural generalization of $K_{Komar}$ that we
desire to find in the present paper.

When the generalized Komar potential $\mathcal{K}$ is established, a formula
for the conserved charges of the Einstein gravity theory admitting
asymptotically AdS spacetime metrics can be proposed as the surface
integral with respect to $\mathcal{K}$, being of the
form\footnote{Throughout the present paper, we take into
account the units in which $G=c=1$.}
\be
\mathcal{Q}=\frac{1}{8\pi} \int_{\partial\Sigma}
\star\mathcal{K}
\, . \label{CCdef}
\ee
As usual, in Eq. (\ref{CCdef}), $\partial\Sigma$ denotes the
$(D-2)$-dimensional $t=const$ surface at spatial infinity ($r=\infty$) if
the conserved charge $\mathcal{Q}$ represents the mass or the angular
momentum, under the coordinate system
$\{t,r,x^i\}$ $(i=1,2,\cdot\cdot\cdot,D-2)$, where $r$ stands for the radial
coordinate. However, when the spacetime metrics are asymptotically flat,
the higher-order derivative term $\Box {\rm d}\xi$ involved in the
potential $\mathcal{K}$ can be excluded. Thus the surface integral
(\ref{CCdef}) turns into the conventional Komar one with a vanishing
cosmological constant. This is attributed to the fact
that the $\Box {\rm d}\xi$ term decreases fast enough so that it vanishes
at spatial infinity.

In order to see the universality of the Komar-like potential (\ref{CalKAdS}),
we move on to comparing it with the ones obtained by other approaches. First,
in comparison with the superpotential $\mathcal{K}^{\mu\nu}_R$ given by
Eq. (35) in \cite{PLnovCC}, here represented by the 2-form
$\mathcal{K}^{\mu\nu}_{gr}$
in the framework of the Einstein gravity armed with the Lagrangian
(\ref{EHLLag}), we follow the work \cite{PLnovCC} (see \cite{PinGR1,PinGR}
as well) to compute
$\mathcal{K}^{\mu\nu}_{gr}$ and find that
\bea
\mathcal{K}^{\mu\nu}_{gr}&=&\frac{c_2}{\ell^2}
\left(R^{\mu\nu}_{\rho\sigma}
-4R^{[\mu}_{[\rho} \delta^{\nu]}_{\sigma]}
\right)\nabla^{\rho}\xi^{\sigma}
+ \frac{(D-2)\ell^2+2c_2R}{2\ell^2}
\nabla^{\mu}\xi^{\nu}  \nn \\
&=&\frac{1}{D-3}\left(\nabla^{[\mu}\xi^{\nu]}
+\frac{1}{2\ell^2}R^{\mu\nu}_{\rho\sigma}\nabla^{\rho}\xi^{\sigma}
\right) \nn \\
&=&\mathcal{K}^{\mu\nu}
\, . \label{PotenforEG}
\eea
It is worth noting that the equation of motion (\ref{Eqmotion}) for the
field has been used in order to obtain the second equality in the above
equation. Second, the potential $\mathcal{K}^{\mu\nu}$ is equivalent with the
KBL superpotential in \cite{KBLSu1,KBLSup} and the potential via the
topological regularization method in even dimensions
\cite{ToPReGR,ACOTZ,Olen4Dker,GMOR,WanPen,GMORTR}.
Third, as what has been shown in \cite{PLnovCC},
the perturbation of  $\mathcal{K}^{\mu\nu}$ about the AdS spacetimes
coincides with the Iyer-Wald potential \cite{IyWald}, the ADT potential
\cite{AbbottD1,AbbottD2,DeserT1,DeserT2,KimKY}, the potential via the
field-theoretical approach \cite{PetLogr}, and the potential given by
\cite{PinGR1,PinGR}. According to these, we further conclude that it is
reasonable to modify the usual Komar potential as the one
$\mathcal{K}^{(3th)}(\xi)$.

What is more, we perform a comparison between the generalized potential
$\mathcal{K}$ and the Komar-like potential put forward within the works
\cite{KasKI,KaRTinJ}, here denoted by $\check{K}^{\mu\nu}$. The potential
$\check{K}^{\mu\nu}$, which can be regarded as the generalization of the
results for the four-dimensional Einstein gravity in \cite{MagKI}, was
constructed out of the equation of motion and the property
$\Box\xi^\mu=-R^\mu_\nu\xi^\nu$ for the divergence-free Killing vector
$\xi^\mu$. It is expressed as the linear combination of the ordinary
Komar potential $K_{Komar}^{\mu\nu}$ with the 2-form Killing potential
$\omega^{\mu\nu}$ defined through the relation
$\xi^\mu=-\nabla_\nu\omega^{\mu\nu}$, that is,
\be
\check{K}^{\mu\nu}=\nabla^{[\mu}\xi^{\nu]}-(D-1)\ell^2\omega^{\mu\nu}
\, . \label{Potomeg}
\ee
Obviously, the potential $\check{K}^{\mu\nu}$ takes a similar structure to
the one $\mathcal{K}^{\mu\nu}$. In fact, we shall see below that both of
them are equivalent at spatial infinity within the scope of the Einstein
gravity. However, an advantage of
$\mathcal{K}^{\mu\nu}$ over $\check{K}^{\mu\nu}$ is that one does not have
to solve the equation $\xi^\mu=-\nabla_\nu\omega^{\mu\nu}$.

With the help of the Bianchi identity
$\nabla_{[\nu}R^{\mu\nu}_{\rho\sigma]}=0$ and the field equation
(\ref{Eqmotion}), we compute the divergence of the skew-symmetric tensor
$R^{\mu\nu}_{\rho\sigma}\nabla^{\rho}\xi^{\sigma}$ in
$\mathcal{K}^{\mu\nu}$, leading to
\bea
\nabla_\nu\big(R^{\mu\nu}_{\rho\sigma}\nabla^{\rho}\xi^{\sigma}\big)&=&
R^{\mu\nu}_{\rho\sigma}\nabla_\nu\nabla^{\rho}\xi^{\sigma}
-2\big(\nabla_{\rho} R^\mu_{\sigma}\big)\nabla^{\rho}\xi^{\sigma}\nn \\
&=&R^{\mu\nu}_{\rho\sigma}R^{\rho\sigma}_{\lambda\nu}\xi^\lambda
\, . \label{CovRxi}
\eea
By utilizing Eqs. (\ref{RiemSI}) and (\ref{CovRxi}), we further obtain
\bea
\frac{1}{2\ell^2}
\nabla_\nu\big(R^{\mu\nu}_{\rho\sigma}\nabla^{\rho}\xi^{\sigma}\big)
\Big|_{\infty}&=&(D-1)\ell^2\xi^\mu \nn \\
&=&-(D-1)\ell^2\nabla_\nu\omega^{\mu\nu}
\, . \label{Comomeg}
\eea
As a consequence of Eq. (\ref{Comomeg}), we conclude that the potentials
$\mathcal{K}^{\mu\nu}$ without the $c_2$ factor and $\check{K}^{\mu\nu}$
coincide with
each other at spatial infinity. Furthermore, as what has been mentioned above,
to completely fix the potential $\check{K}^{\mu\nu}$, a necessary procedure
is to solve the Killing potential $\omega^{\mu\nu}$ from the equation
$\xi^\mu=-\nabla_\nu\omega^{\mu\nu}$, which is of great difficulty to handle,
particularly for rotating spacetimes in higher dimensions. As a solution,
we propose that the 2-form
$R^{\mu\nu}_{\rho\sigma}\nabla^{\rho}\xi^{\sigma}$ can be a proper substitute
for the Killing potential $\omega^{\mu\nu}$ involved in the potential
$\check{K}^{\mu\nu}$, that is,
\be
\omega^{\mu\nu}\rightarrow-\frac{1}{2(D-1)\ell^4}
R^{\mu\nu}_{\rho\sigma}\nabla^{\rho}\xi^{\sigma}
\, . \label{Omegeq}
\ee

%%%%%%%%%%%%%%%%%%%%%%%%%%%%%%%%%%%%%%%%%%%%%%%%%%%%%%%%%%%%%%%%%%%%%%%%
\section{Applications in black holes with an AdS asymptotic}\label{three}
%%%%%%%%%%%%%%%%%%%%%%%%%%%%%%%%%%%%%%%%%%%%%%%%%%%%%%%%%%%%%%%%%%%%%%%%

In this section, we are going to focus on applications of the Komar-like
integral (\ref{CCdef}) in the computations for conserved charges of various
asymptotically AdS black holes, including the $D$-dimensional
Schwarzschild-AdS black hole, the four-dimensional regular AdS black hole,
the four-dimensional asymptotically AdS Kerr-Sen black hole together with
its ultra-spinning generalization, the four-dimensional Kerr-NUT-AdS black
hole, and the Kerr-AdS black hole in arbitrary dimensions. By contrast with
other existing approaches, the calculations are much simpler. All the results
support that the surface integral in terms of the improved Komar potential
with higher-order corrections indeed yield the physical mass and angular
momentum in a unified way
within the scope of the Einstein gravity in the presence of the
negative cosmological constant.

%%%%%%%%%%%%%%%%%%%%%%%%%%%%%%%%%%%%%%%%%%%%%%%%%%%%%%%%%%%%%%%%%%%%%%%%
\subsection{Mass of Schwarzschild-AdS black holes in arbitrary dimensions
and four-dimensional regular AdS black holes}
\label{three1}
%%%%%%%%%%%%%%%%%%%%%%%%%%%%%%%%%%%%%%%%%%%%%%%%%%%%%%%%%%%%%%%%%%%%%%%%

In this subsection, on basis of the formula (\ref{CCdef}) for conserved
charges, we will calculate the mass of $D$-dimensional
Schwarzschild-AdS black holes, as well as four-dimensional spherically
symmetric regular AdS black holes.

Without loss of generality, let us take into consideration the static
Schwarzschild-like metric ansatz being of the form
\be
ds^2=-F(r)dt^2+\frac{dr^2}{F(r)X(r)}+Y^2(r)h_{ij}\big(x^k\big)dx^idx^j
\, , \label{sleDD}
\ee
where $h_{ij}$ is the metric for $(D-2)$-dimensional space. Our goal is to
apply the formula (\ref{CCdef}) to compute the mass of the solutions equipped
with the metric ansatz (\ref{sleDD}). The associated Killing vector is chosen as
$\xi^\mu=-\delta^\mu_t$. Under this to evaluate the $(t,r)$ component for
the covariant derivative with respect to the vector $\xi^\mu$,  we have
\be
\nabla^t\xi^r=-\nabla^r\xi^t=\frac{1}{2}XF^\prime
\, . \label{CovXitr}
\ee
Here and in the remainder of this subsection, the function with prime denotes
its derivative with respect to the radial coordinate $r$.
The $(t,r,\rho,\sigma)$ component of the Riemann curvature tensor
$R^{\mu\nu}_{\rho\sigma}$ is read off as
\be
R^{tr}_{\rho\sigma}=-R^{rt}_{\rho\sigma}
=-\frac{1}{2}\left(F^{\prime}X^{\prime}
+2XF^{\prime\prime}\right)
\delta^t_{[\rho}\delta^r_{\sigma]}
\, . \label{Rtrrhsig}
\ee
Substituting Eqs. (\ref{CovXitr}) and (\ref{Rtrrhsig}) into the Komar-like
potential $\mathcal{K}^{\mu\nu}$ in Eq. (\ref{CalKAdS}), we obtain its
$(t,r)$ component of the form
\be
\mathcal{K}^{tr}=c_2XF^{\prime}\left(1
-\frac{XF^{\prime\prime}}{2\ell^2}
-\frac{F^{\prime}X^\prime}{4\ell^2}\right)
\, . \label{KtrStS}
\ee
As a consequence, under the formula (\ref{CCdef}) for the conserved charges,
the mass is defined through the integral of $\mathcal{K}^{tr}$ over the
$(D-2)$-dimensional space with the metric $h_{ij}$, presented by
\be
M=\frac{c_2V_{D-2}}{8\pi}\lim_{r\rightarrow\infty}
F^{\prime}|Y|^{D-2}\sqrt{|X|}\left(1
-\frac{XF^{\prime\prime}}{2\ell^2}
-\frac{F^{\prime}X^\prime}{4\ell^2}\right)
\, , \label{ConCofStS}
\ee
where the volume for the $(D-2)$-dimensional space
$V_{D-2}=\int\sqrt{h}dx^{D-2}$ with the determinant of the metric tensor
$h=det(h_{ij})$.

For concreteness, we take into account the mass of the $D$-dimensional
Schwarzschild-AdS black holes. In such a case, all the undetermined
quantities in the line element (\ref{sleDD}) are given by
\bea
F(r)&=&1-\frac{2m}{r^{D-3}}+\ell^2r^2\, , \quad X(r)=1 \, ,\nn \\
Y(r)&=&r \, , \quad d\Omega_{D-2}^2=h_{ij}dx^idx^j\, .
\label{SchAdS}
\eea
In the above equation, the integration constant $m$ is related to the mass,
and $d\Omega_{D-2}^2$ is the line element on a $(D-2)$-dimensional unit
sphere. By making use of Eq. (\ref{ConCofStS}), we obtain the mass of the
Schwarzschild-AdS black holes $M_{SAdS}$, being of the form
\be
M_{SAdS}=\frac{mc_2(D-2)(D-3)V_{D-2}}{4\pi}
\, . \label{MassSchAdS}
\ee
Here the volume of the $(D-2)$-dimensional unit sphere $V_{D-2}$ is given by
\be
V_{D-2}=\frac{2\pi^{(D-1)/2}}{\Gamma\big(\frac{D-1}{2}\big)}
\, . \label{VUnitspher}
\ee
When the $c_2$ parameter is given by Eq. (\ref{kval2}),
the mass for the Schwarzschild-AdS black holes $M_{SAdS}$ takes the same
value as that in the existing literature
\cite{AMDmass2,BarnichC,DasMCT,GibPP,DerKather,GullTek,OleKou,JinPen}.
Besides, it returns to the standard result for the mass of the
Schwarzschild black holes when $\ell=0$. In other words,
these confirm the value of $c_2$. On the other hand, if the
third-order derivatives in $M$ are neglected, it is observed from
Eq. (\ref{ConCofStS}) that $M_{SAdS}$ diverges at infinity since
$\big(r^{D-2}F^{\prime}\big)\big|\rightarrow\infty$.
This demonstrates that the higher-order derivative terms just cancel out the
divergent ones existing in the original Komar potential.

As another application to the static spacetime, one is able to adopt
straightforwardly the formula (\ref{ConCofStS}) to
calculate the mass of the so-called Bardeen-type regular AdS black holes
constructed in the works \cite{MTKreA,FWreAdS}. For generality, we consider
the metric in \cite{FWreAdS}, given by
\be
ds^2=-f(r)dt^2+\frac{dr^2}{f(r)}+r^2\big(d\theta^2+\sin^2\theta d\phi^2\big)
\, , \label{ELregAdS}
\ee
with the function
\be
f(r)=1-\frac{2m}{r}-\frac{2q^3r^{k}}{\alpha r\big(r^l+q^l\big)^{k/l}}
+\ell^2r^2\, .
\ee
Here, $m$, $q$, $\alpha$, $k$ and $l$ are constant parameters. The direct
calculations on base of Eq. (\ref{ConCofStS}) give rise to the mass
$M_{RegAdS}$ for the asymptotically AdS regular black holes, presented by
means of
\be
M_{RegAdS}=m+\frac{q^3}{\alpha}
\, . \label{MregABH}
\ee
$M_{RegAdS}$ coincides with the AMD mass in \cite{FWreAdS}. Obviously, the
inclusion of the $q^3/\alpha$ term modifies the mass of the four-dimensional
Schwarzschild-AdS black holes.

%%%%%%%%%%%%%%%%%%%%%%%%%%%%%%%%%%%%%%%%%%%%%%%%%%%%%%%%%%%%%%%%%%%%%%%%
\subsection{Mass and angular momentum for Kerr-Sen-AdS$_4$ black holes}
\label{three2}
%%%%%%%%%%%%%%%%%%%%%%%%%%%%%%%%%%%%%%%%%%%%%%%%%%%%%%%%%%%%%%%%%%%%%%%%

In this subsection, we are going to compute the mass and angular momentum
for the four-dimensional asymptotically AdS Kerr-Sen (Kerr-Sen-AdS$_4$)
black holes given by \cite{KSAdS4}. We adopt the line element in a
non-rotating frame at infinity, taking the following form in
Boyer-Lindquist coordinates
\bea
ds^2&=& -\frac{\Delta_r}{\Sigma}\left[dt-\Upsilon_{\theta}
\left(\frac{d\phi}{\Xi} - a\ell^2 \frac{dt}{\Xi}\right)\right]^2
+\frac{\Sigma}{\Delta_{r}}dr^2+\frac{\Sigma}{\Delta_{\theta}}
d\theta^2\nn \\
&&+\frac{\Delta_{\theta}\sin^2\theta}{\Sigma}\left[adt-\Upsilon_{r}
\left(\frac{d\phi}{\Xi} - a\ell^2\frac{dt}{\Xi}\right)\right]^2
\, . \label{4DKeSAdS}
\eea
In Eq. (\ref{4DKeSAdS}), the functions $\Upsilon_{\theta}$,
$\Upsilon_{r}$, $\Delta_{\theta}$, $\Sigma$, and $\Delta_{r}$ are given by
\bea
\Upsilon_{\theta}&=&a\sin^2\theta \, , \nn \\
\Upsilon_{r}&=&r^2+2br+a^2 \, ,  \nn \\
\Delta_{\theta}&=&1-a^2\ell^2\cos^2\theta \, , \nn \\
\Sigma&=&r^2+2br+a^2\cos^2\theta \, , \nn \\
\Delta_{r}&=&\Upsilon_{r}[1+\ell^2(r^2+2br)]-2mr
\, , \label{Func4DKSAdS}
\eea
respectively. the constant parameter $\Xi = 1-\ell^2 a^2$, and $(m,a,b)$
are integration constants. When $b=0$, the metric ansatz (\ref{4DKeSAdS})
turns into that of the four-dimensional Kerr-AdS black hole solution
\cite{4DKeAdSsolu}.

The mass $M_{KSAdS}$ and the angular momentum $J_{KSAdS}$ for the
Kerr-Sen-AdS$_4$ black hole are associated
with the Killing vectors $\xi^\mu_{(t)}=(-1,0,0,0)$ and
$\xi^\mu_{(\phi)}=(0,0,0,1)$, respectively. In this setting, on basis of
Eq. (\ref{CalKAdS}) to evaluate the $(t,r)$ component of the Komar-like
potential, we arrive at
\bea
\sqrt{-g}\mathcal{K}^{tr}\big(\xi^\mu_{(t)}\big)&=&
\frac{m\left(3\Delta_{\theta}-\Xi\right)\sin\theta}{\Xi^2}+O\left(\frac{1}{r}\right)
\, , \nn \\
\sqrt{-g}\mathcal{K}^{tr}\big(\xi^\mu_{(\phi)}\big)&=&
\frac{3m\sin\theta\Upsilon_{\theta}}{\Xi^2}+O\left(\frac{1}{r}\right)
\, , \label{KtrforKSAdS}
\eea
where $\sqrt{-g}=(\Upsilon_{r}-a\Upsilon_{\theta})\sin\theta/\Xi$.
The integration of Eq. (\ref{KtrforKSAdS}) with respect to the coordinate
$\theta$ further leads to
\be
M_{KSAdS}=\frac{m}{\Xi^2} \, , \quad
J_{KSAdS}=\frac{ma}{\Xi^2}
\, .\label{MJofKSAdS}
\ee
On the other hand, like in \cite{KSAdS4}, if one computes the mass and
angular momentum of the Kerr-Sen-AdS$_4$ black hole in the rotating frame,
that is, $\phi\rightarrow \phi+a\ell^2t$ in the metric (\ref{4DKeSAdS}),
the mass $M_{KSAdS}$ becomes $m/\Xi$, while $J_{KSAdS}$ remains the same
value.

It is worth noting that the scalar fields and the U(1) gauge field are
not involved in the computations for the conserved charges of the
Kerr-Sen-AdS$_4$ black hole. As a matter of fact, one can follow the
work \cite{PenheSt} to confirm that those fields decrease fast enough at
asymptotic infinity so that they make no contribution to the charges.
This holds true to the four-dimensional Kerr-Newman-AdS black holes
\cite{4DKeAdSsolu,4DKNUTC}. Therefore, their mass and angular momentum
are consistent with the ones in Eq. (\ref{MJofKSAdS}).

What is more, in terms of the metric given by Eq. (13) in \cite{KSAdS4},
one makes use of the generalized Komar potential (\ref{CalKAdS}) with
$\ell$ substituted by $l^{-1}$ to compute the
mass and the angular momentum of the ultra-spinning Kerr-Sen-AdS$_4$
black hole. One can obtain the mass $M=\mu m/(2\pi)$ and the angular
momentum $J=\mu ml/(2\pi)$ in \cite{KSAdS4}.

%%%%%%%%%%%%%%%%%%%%%%%%%%%%%%%%%%%%%%%%%%%%%%%%%%%%%%%%%%%%%%%%%%%%%%%%
\subsection{Conserved quantities of four-dimensional
Kerr-NUT-AdS black holes}\label{three3}
%%%%%%%%%%%%%%%%%%%%%%%%%%%%%%%%%%%%%%%%%%%%%%%%%%%%%%%%%%%%%%%%%%%%%%%%

The metric ansatz of the four-dimensional Kerr-NUT-AdS black holes, which
has been known for some time \cite{4DKNUTC,4DKNuPK,CLPKNuT}, can be written
as the same form as Eq. (\ref{4DKeSAdS}). However, the functions
$\big(\Upsilon_{r},\Delta_{r},\Delta_{\theta},\Upsilon_{\theta},\Sigma\big)$
are replaced with the ones
$\big(\hat{\Upsilon}_{r},\hat{\Delta}_{r},\hat{\Delta}_{\theta},
\hat{\Upsilon}_{\theta},\hat{\Sigma}\big)$, respectively. Here the functions
$\hat{\Upsilon}_{r}$ and $\hat{\Delta}_{r}$ with respect to the coordinate
$r$ are presented by
\bea
\hat{\Upsilon}_{r}&=&r^2+(a+n)^2 \, , \nn\\
\hat{\Delta}_{r}&=&r^2+\ell^2r^2(r^2+6n^2+a^2)-2mr \nn \\
&&+(3\ell^2n^2+1)(a^2-n^2)
\, , \label{Func4DKNUTAdS}
\eea
while the functions $\big(\hat{\Delta}_{\theta},
\hat{\Upsilon}_{\theta},\hat{\Sigma}\big)$ are given by
\bea
\hat{\Delta}_{\theta}&=&1-a\ell^2\cos\theta(4n+a\cos\theta) \, , \nn \\
\hat{\Upsilon}_{\theta}&=&a\sin^2\theta+2n(1-\cos\theta) \, , \nn \\
\hat{\Sigma}&=&r^2+(n+a\cos\theta)^2
\, . \label{Func4DKNUTAdS2}
\eea
In Eqs. (\ref{Func4DKNUTAdS}) and (\ref{Func4DKNUTAdS2}), $n$ is the NUT
charge parameter. Accordingly, the $(t,r)$ components of the potential
$\mathcal{K}^{\mu\nu}$ related to the Killing vectors $\xi^\mu_{(t)}$ and
$\xi^\mu_{(\phi)}$ are read off as
\bea
\sqrt{-g}\hat{\mathcal{K}}^{tr}\big(\xi^\mu_{(t)}\big)&=&
\frac{m\left(3a\ell^2\hat{\Upsilon}_{\theta}+2\Xi\right)\sin\theta}{\Xi^2}
+O\left(\frac{1}{r}\right)
\, , \nn \\
\sqrt{-g}\hat{\mathcal{K}}^{tr}\big(\xi^\mu_{(\phi)}\big)&=&
\frac{3m\sin\theta\hat{\Upsilon}_{\theta}}{\Xi^2}+O\left(\frac{1}{r}\right)
\, . \label{KtrforKNUTAdS}
\eea
Furthermore, in terms of the formulation (\ref{CCdef}) for conserved
quantities to compute the mass $M_{KNUT}$ and the angular momentum
$J_{KNUT}$ for the four-dimensional Kerr-NUT-AdS black hole, we have
\be
M_{KNUT}= \frac{m(1+3an\ell^2)}{\Xi^2}\, , \quad
J_{KNUT}=\frac{m(a+3n)}{\Xi^2}
\, . \label{MJKNUT}
\ee

Letting the NUT parameter $n$ in Eq. (\ref{MJKNUT}) vanish, we see that
the mass $M_{KNUT}$ and the angular momentum $J_{KNUT}$ becomes the ones of
the four-dimensional Kerr-AdS black hole, respectively. Although here we
only consider the computations of the mass and the angular momentum for
the four-dimensional Kerr-NUT-AdS black holes, we expect that the
formula (\ref{CCdef}) are applicable to the higher-dimensional ones. This
will be verified in the future work.

%%%%%%%%%%%%%%%%%%%%%%%%%%%%%%%%%%%%%%%%%%%%%%%%%%%%%%%%%%%%%%%%%%%%%%%%
\subsection{Conserved charges for $D$-dimensional
Kerr-AdS black holes}\label{three4}
%%%%%%%%%%%%%%%%%%%%%%%%%%%%%%%%%%%%%%%%%%%%%%%%%%%%%%%%%%%%%%%%%%%%%%%%

In the present subsection, the mass and angular momenta for the generic
$D$-dimensional Kerr-AdS black holes constructed in \cite{GLuPP1,GLuPP2}
will be computed in terms of the formula (\ref{CCdef}). As is known,
these black holes can be seen as the higher-dimensional generalizations
of the four- and five-dimensional Kerr-AdS black holes given by
\cite{4DKeAdSsolu,5DsoluHHT}, as well as the extensions including a
cosmological constant to the asymptotically flat Myers-Perry black holes
in $D$ dimensions \cite{MyersP}.

The metric tensors for the $D$-dimensional Kerr-AdS black holes satisfy the
field equation (\ref{Eqmotion}). They possess $n=(D-\varepsilon-1)/2$
($\varepsilon=1$ for $D$ even and $\varepsilon=0$ for $D$ odd) independent
rotations in $n$ orthogonal 2-planes, characterized by $n$ parameters
$a_i$ $(1\leq i\leq n)$ and $n$ azimuthal angles $\phi_i$, which are
$2\pi$-periodic. In the coordinate system
$\{t,r,\mu_1,\cdot\cdot\cdot,\mu_{n+\varepsilon-1},
\phi_1,\cdot\cdot\cdot,\phi_n\}$,
the line element of the spacetime for the $D$-dimensional Kerr-AdS black
hole is of the form \cite{GLuPP1,GLuPP2}
\be
ds^2_{(D)}=d\bar{s}^2_{(D)}+\frac{2mr^2U}{H(V-2m)}dr^2
+\frac{2mH}{r^2UV}
\left(Wdt-\sum_{i=1}^{n}a_i\mu_i^2\frac{d\phi_i}{\Xi_i}\right)^2
\, , \label{DDKerrAdS}
\ee
in which the line element $d\bar{s}^2_{(D)}$ is read off as
\bea
d\bar{s}^2_{(D)}&=&-HWdt^2
+\sum_{i=1}^{n}\mu_i^2(r^2+a_i^2)\frac{d\phi_i^2}{\Xi_i}
+\sum_{i=1}^{n+\varepsilon}(r^2+a_i^2)\frac{d\mu_i^2}{\Xi_i}
\nn \\
&&-\frac{\ell^2}{HW}
\left(\sum_{i=1}^{n+\varepsilon}\frac{r^2+a_i^2}{\Xi_i}\mu_id\mu_i\right)^2
+\frac{r^2U}{H}dr^2
\, . \label{DbarsDK}
\eea
In Eqs. (\ref{DDKerrAdS}) and (\ref{DbarsDK}), the constant parameters
$\Xi_i=1-a_i^2\ell^2$ $(1\leq i\leq n)$, while $\Xi_{n+1}=1$, arising from
that $a_{n+1}=0$, when $D$ is even. The four functions $(H,U,V,W)$ are
presented respectively by
\bea
H&=&1+\ell^2r^2\, , \quad
U=\sum_{i=1}^{n+\varepsilon}\frac{\mu_i^2}{r^2+a_i^2}
\, , \nn \\
V&=&r^{\varepsilon-2}H\prod_{i=1}^n(r^2+a_i^2)
\, , \nn \\
W&=&\sum_{i=1}^{n+\varepsilon}\frac{\mu_i^2}{\Xi_i}
\, . \label{WUVdphi}
\eea
The $\mu_i$ variables are constrained by
$\sum_{i=1}^{n+\varepsilon}\mu_i^2=1$, from which $\mu_{n+\varepsilon}$
can be solved as
$|\mu_{n+\varepsilon}|=\sqrt{1-\sum_{k=1}^{n+\varepsilon-1}\mu_k^2}$. Under such
a choice of $\mu_{n+\varepsilon}$, each $\mu_i$ is assumed to range over
$0\leq \mu_i\leq 1$ $(i=1,2,\cdot\cdot\cdot,n)$, while the $\mu_{n+1}$ variable
runs from -1 to 1 in even $D$ case. The value of
$\sqrt{-g}$ is written as \cite{{GibPP}}
\be
\sqrt{-g}=\frac{r^3UV\prod_{j=1}^n\mu_j}{H\mu_{n+\varepsilon}\prod_{i=1}^n\Xi_i}
\, . \label{sqrtg}
\ee

For the Kerr-AdS black hole (\ref{DDKerrAdS}), the Killing vectors associated
with the mass $M_{KAdS}$ and the angular momentum $J^{(i)}_{KAdS}$
along the $\phi_i$ direction are chosen as $\xi^\mu_{(t)}=-\delta^\mu_t$
and $\xi^\mu_{(\phi_i)}=\delta^\mu_{\phi_i}$, respectively. By utilizing
Eq. (\ref{CalKAdS}) for the superpotential $\mathcal{K}^{\mu\nu}$, we
have
\bea
\sqrt{-g}\mathcal{K}^{tr}_{KAdS}\big(\xi^\mu_{(t)}\big)&=&
\frac{m[(D-1)W-1]\prod_{j=1}^n\mu_j}{\mu_{n+\varepsilon}\prod_{k=1}^n\Xi_k}
+O\left(\frac{1}{r}\right)
\, , \nn \\
\sqrt{-g}\mathcal{K}^{tr}_{KAdS}\big(\xi^\mu_{(\phi_i)}\big)&=&
\frac{ma_i(D-1)\mu_i^2\prod_{j=1}^n\mu_j}{\mu_{n+\varepsilon}\Xi_i\prod_{k=1}^n\Xi_k}
+O\left(\frac{1}{r}\right)
\, . \label{KtrforKAdS}
\eea
The results in Eq. (\ref{KtrforKAdS}) were also obtained by using the KBL
superpotential method in \cite{DerKather}, as well as by means of the BBC
formalism in \cite{BarnichC}.
Substituting Eq. (\ref{KtrforKAdS}) into the Komar-like integral (\ref{CCdef})
to calculate the mass $M_{KAdS}$ and the angular momentum $J^{(i)}_{KAdS}$,
one obtains the following results:
\bea
M_{KAdS}&=&\frac{V_{D-2}}{4\pi}\frac{m}{\prod_{j=1}^n\Xi_j}
\left(\sum_{i=1}^{n}\frac{1}{\Xi_i}-\frac{1-\epsilon}{2}\right)
\, , \nn \\
J^{(i)}_{KAdS}&=&\frac{V_{D-2}}{4\pi}\frac{ma_i}{\Xi_i\prod_{j=1}^n\Xi_j}
\, . \label{MassAninDdi}
\eea
Here the volume of the $(D-2)$-dimensional sphere $V_{D-2}$ is presented
in (\ref{VUnitspher}).
It should be pointed out that the following identity has
been used in the calculations
\be
\int \frac{\mu_i^2\prod_{j=1}^n\mu_j}{\mu_{n+\varepsilon}}
\left(\prod_{k=1}^{n+\varepsilon-1}d\mu_k\right)
\prod_{l=1}^{n}d\phi_l =\frac{2V_{D-2}}{D-1}
\, , \label{Intemuphi}
\ee
when the integer $i$ ranges over $1\leq i \leq n$, together with the
following integral derived from Eq. (\ref{Intemuphi})
\be
\int \frac{\mu_{n+1}^2\prod_{j=1}^n\mu_j}{\mu_{n+1}}
\left(\prod_{k=1}^{n}d\mu_k\right)
\prod_{l=1}^{n}d\phi_l =\frac{V_{D-2}}{D-1}
\,  \label{InteevenD}
\ee
in even $D$ case. What is more, due to that
$\big(\Box{\rm d}\xi_{(\phi_i)}\big)^{tr}$ multiplied by the factor
$\sqrt{-g}$ vanishes at infinity, the third-order derivative term
in the Komar-like potential (\ref{CalKAdS}) makes no contributions to
the angular momentum. In this sense, the usual Komar integral is enough
to yield the angular momentum of $D$-dimensional Kerr-AdS black holes
\cite{GibPP}.

When all the rotation parameters $a_i$'s vanish, $M_{KAdS}$ coincides with
the mass $M_{SAdS}$ for the $D$-dimensional Schwarzschild-AdS black holes
given by Eq. (\ref{MassSchAdS}). The mass $M_{KAdS}$ and the angular
momentum $J^{(i)}_{KAdS}$ in Eq. (\ref{MassAninDdi}) are congruent with
the results via other methods in the literature
\cite{Olen4Dker,BarnichC,DasMCT,GibPP,DerKather,GullTek,OleKou,JinPen}.

%%%%%%%%%%%%%%%%%%%%%%%%%%%%%%%%%%%%%%%%%%%%%%%%%%%%%%%%%%%%%%%%%%%%%%%%
\section{Summary}\label{four}
%%%%%%%%%%%%%%%%%%%%%%%%%%%%%%%%%%%%%%%%%%%%%%%%%%%%%%%%%%%%%%%%%%%%%%%%

Within this paper, for the sake of generalizing the usual Komar integral
to the Einstein gravity theory equipped with the Lagrangian (\ref{EHLLag}),
we modify the usual Komar current as the identically conserved
one $\mathcal{J}^{(4th)}$ given by Eq. (\ref{CalJ4th}). This extended
current being comprised of second- and fourth-order derivative terms of the
Killing vector field $\xi$ is generated via the action of the three
differential operators $\big(\Box,{\rm d},\hat{\delta}\big)$ on $\xi$.
Through the well-known relationship between the conserved
current and potential, we then derive the potential $\mathcal{K}^{(3th)}$
in Eq. (\ref{CalK3th}). Such a potential can be interpreted as
the higher-order derivative generalization of the usual Komar potential,
due to the fact that it contains third-order derivative terms
${\rm d}\Box\xi$ and $\Box{\rm d}\xi$ apart from the Komar potential
${\rm d}\xi$. Particularly, by applying the potential with higher-order
corrections to the usual Komar one into the Einstein gravity with a
negative cosmological constant, we obtain the Komar-like potential
$\mathcal{K}$ in Eq. (\ref{CalKAdS}), which is equivalent with some
potentials in the literature. The surface integral with respect to the
potential $\mathcal{K}^{\mu\nu}$ further gives rise to a unified definition
(\ref{CCdef}) for the mass and the angular momentum of spacetimes with an
AdS asymptotic. As applications of the formula (\ref{CCdef}), we compute the
mass and the angular momentum of some asymptotically AdS black holes, such
as the $D$-dimensional Schwarzschild-AdS black hole, the Bardeen-type
regular AdS black hole, the four-dimensional asymptotically AdS Kerr-Sen
black hole, the Kerr-NUT-AdS black hole, and the Kerr-AdS black holes in
arbitrary dimensions. All the results are consistent with the ones via other
methods.

There are several issues associated with the present paper which should
deserve further investigation. Firstly, as is known, the usual Komar current
and potential can be derived from the variation of the
Einstein-Hilbert-$\Lambda$ Lagrangian under the guidance of Noether theorem.
Thus, in parallel with this, it is better to find an effective
Lagrangian, from which the current $\mathcal{J}^{(4th)}$ and the potential
$\mathcal{K}^{(3th)}$ are able to be derived by following the standard
Noether method. Secondly, it is of great interest to extend the higher-order
corrected potential $\mathcal{K}^{(3th)}$ to investigate the defintions of
the conserved charges for spacetimes with asymptotical structures different
from the asymptotically AdS one. Thirdly, to see further the universality of
the Komar-like potential $\mathcal{K}$, it can be utilized to understand the
thermodynamical properties of various spacetimes with an AdS asymptotic,
such as the derivation of the Smarr formula for the first law and enthalpy.

\section*{Acknowledgments}

This work was supported by the Natural Science Foundation of
China under Grant Nos. 11865006 and 11505036. It was also partially supported
by the Technology Department of Guizhou province Fund
under Grant Nos. [2018]5769.

%%%%%%%%%%%%%%%%%%%%%%%%%%%%%%%%%%%%%%%%%%%%%%%%%%%%%%%%%%%
\appendix
%%%%%%%%%%%%%%%%%%%%%%%%%%%%%%%%%%%%%%%%%%%%%%%%%%%%%%%%
\section{Inclusions of more than fourth-order derivatives within
conserved currents} \label{appendA}
%%%%%%%%%%%%%%%%%%%%%%%%%%%%%%%%%%%%%%%%%%%%%%%%%%%%

According to the rules in the work \cite{PenZ}, all the currents generated
under the action of the $(\Box,{\rm d},\hat{\delta})$ operators on an
arbitrary Killing vector contain only even-order derivatives of this vector.
As a consequence, except for the
derivative of order two, the lowest differential order derivative of the
Killing vector is the fourth-order one. To this point, for simplicity, one
could give priority to the fourth-order derivatives of the Killing vector
when the higher-order corrections to the ordinary Komar current are under
consideration. In principle, if these are inadequate to meet the requirement
for the convergence of conserved charges, the sixth- or higher-order ones can
be continuously considered to be incorporated into the conserved currents.
However, in this appendix, it will be demonstrated that the fourth-order
derivatives are sufficient within the context of the Einstein gravity theory
in the presence of a negative cosmological constant.

We first take into consideration of the inclusion of sixth-order derivative
terms of the Killing vector $\xi$. In this general case, the most generic 1-form
$J^{(6th)}_{gen}(\xi)$, generated by the action of the
$(\Box,{\rm d},\hat{\delta})$ operators upon $\xi$, is decomposed as
\be
J^{(6th)}_{gen}(\xi)=J^{(6th)}_{\Box}(\xi)
+J^{(6th)}_{{\rm d}}(\xi)
+J^{(6th)}_{\hat{\delta}}(\xi)
\, . \label{J6thgen}
\ee
Here the 1-form $J^{(6th)}_{\Box}(\xi)$ is presented by
\bea
J^{(6th)}_{\Box}(\xi)
&=&\alpha_{31}\Box\hat{\delta}{\rm d}\hat{\delta}{\rm d}\xi
+\alpha_{32}\Box^2\hat{\delta}{\rm d}\xi
+\alpha_{33}\Box\hat{\delta}\Box{\rm d}\xi\nn \\
&&+\alpha_{34}\Box\hat{\delta}{\rm d}\Box\xi
+\alpha_{35}\Box{\rm d}\hat{\delta}\Box\xi
+\alpha_{36}\Box^3\xi
\, , \label{Jbox6th}
\eea
the closed 1-form $J^{(6th)}_{{\rm d}}(\xi)$ is of the form
\be
J^{(6th)}_{{\rm d}}(\xi)
=\beta_{31}{\rm d}\hat{\delta}\Box\hat{\delta}{\rm d}\xi
+\beta_{32}{\rm d}\Box\hat{\delta}\Box\xi
+\beta_{33}{\rm d}\hat{\delta}\Box\Box\xi
\, , \label{Jd6th}
\ee
and the identically conserved $J^{(6th)}_{\hat{\delta}}(\xi)$ is given by
\bea
J^{(6th)}_{\hat{\delta}}(\xi)
&=&\lambda_{31}\big(\hat{\delta}{\rm d}\big)^3\xi
+\lambda_{32}\hat{\delta}\Box{\rm d}\hat{\delta}{\rm d}\xi
+\lambda_{33}\hat{\delta}{\rm d}\Box\hat{\delta}{\rm d}\xi\nn \\
&&+\lambda_{34}\hat{\delta}{\rm d}\hat{\delta}\Box{\rm d}\xi
+\lambda_{35}\hat{\delta}{\rm d}\Box^2\xi
+\lambda_{36}\hat{\delta}\Box^2{\rm d}\xi \nn \\
&&+\lambda_{37}\hat{\delta}\Box{\rm d}\Box\xi
+\lambda_{38}\hat{\delta}{\rm d}\hat{\delta}{\rm d}\Box\xi
\, . \label{Jhd6th}
\eea
In Eqs. (\ref{Jbox6th}), (\ref{Jd6th}) and (\ref{Jhd6th}), $\alpha_{3i}$'s,
$\beta_{3i}$'s and $\lambda_{3i}$'s represent constant parameters. For a
particular Killing vector satisfying
$\hat{\delta}J^{(6th)}_{\Box}(\xi)
=-\hat{\delta}J^{(6th)}_{{\rm d}}(\xi)$, the 1-form $J^{(6th)}_{gen}(\xi)$
is conserved. However, in order to guarantee that
$\hat{\delta}J^{(6th)}_{gen}(\xi)=0$ holds identically for any Killing vector,
it is required that both the 1-forms $J^{(6th)}_{\Box}(\xi)$ and
$J^{(6th)}_{{\rm d}}(\xi)$ vanish. As a consequence, we obtain the identically conserved
current comprised of the sixth-order derivatives of the Killing vector
$J^{(6th)}(\xi)=J^{(6th)}_{\hat{\delta}}(\xi)$. For convenience, making
use of the identity $\hat{\delta}{\rm d}\xi=2\Box\xi$ to reformulate the
current $J^{(6th)}(\xi)$ results in
\bea
J^{(6th)}(\xi)&=&
-k_{61}\hat{\delta}\Box{\rm d}\Box\xi
-k_{62}\hat{\delta}{\rm d}\Box\Box\xi
-k_{63}\hat{\delta}{\rm d}\hat{\delta}{\rm d}\Box\xi \nn \\
&&-k_{64}\hat{\delta}{\rm d}\hat{\delta}\Box{\rm d}\xi
-k_{65}\hat{\delta}\Box\Box{\rm d}\xi
\, , \label{CalJ6th}
\eea
where $k_{6i}$'s denote arbitrary constant parameters. The relationship
$J^{(6th)}(\xi)=-\hat{\delta}K^{(5th)}(\xi)$
gives rise to the potential
\bea
K^{(5th)}(\xi)&=&
k_{61}\Box{\rm d}\Box\xi
+k_{62}{\rm d}\Box\Box\xi
+k_{63}{\rm d}\hat{\delta}{\rm d}\Box\xi \nn \\
&&+k_{64}{\rm d}\hat{\delta}\Box{\rm d}\xi
+k_{65}\Box\Box{\rm d}\xi
\, . \label{CalK5th}
\eea

Next, as a particular case of Eq. (\ref{CalK5th}), we switch to dealing
with the potentials within the framework of the Einstein gravity theory
endowed with the Lagrangian (\ref{EHLLag}). Under the identity
$\Box\xi=(D-1)\ell^2\xi$, the potential $K^{(5th)}(\xi)$ turns into
\bea
K^{(5th)}_{gr}(\xi)&=&
k_{61}(D-1)\ell^2\Box{\rm d}\xi
+(k_{62}+2k_{63})(D-1)^2\ell^4{\rm d}\xi \nn \\
&&+k_{64}{\rm d}\hat{\delta}\Box{\rm d}\xi
+k_{65}\Box\Box{\rm d}\xi
\, . \label{CalK5gr}
\eea
Apparently, the terms proportional to $\Box{\rm d}\xi$ and ${\rm d}\xi$
in $K^{(5th)}_{gr}(\xi)$ can be covered by $\mathcal{K}^{(3th)}(\xi)$
in (\ref{CalK3th}). What is more, due to that
\be
{\rm d}\hat{\delta}\Box{\rm d}\xi=
-2\Big(R_{\alpha\beta}^{\lambda\sigma}R^{\alpha\beta}_{\lambda\mu}
\nabla_{\nu}\xi_{\sigma}
+\xi_{\alpha}R^{\rho\sigma}_{\lambda\mu}
\nabla_{\nu}R_{\rho\sigma}^{\lambda\alpha}\Big)
{\rm d}x^\mu\wedge{\rm d}x^\nu
\, ,
\ee
by making use of the falloff condition (\ref{RiemSI}) for the Riemann
curvature tensor and the equation
$\big(R_{\rho\sigma}^{\lambda[\mu}
\nabla^{\nu]}R^{\rho\sigma}_{\lambda\alpha}\big)\big|_{\infty}=0$, we
observe that
$\big({\rm d}\hat{\delta}\Box{\rm d}\xi\big)\big|_{\infty}
=4(D-1)\ell^4{\rm d}\xi$, proportional to ${\rm d}\xi$. This implies that
${\rm d}\hat{\delta}\Box{\rm d}\xi$ overlaps ${\rm d}\xi$ in
$\mathcal{K}^{(3th)}(\xi)$ at infinity. Therefore, the linear combination of
$\mathcal{K}^{(3th)}(\xi)$ with $K^{(5th)}_{gr}(\xi)$ can be equivalently
described by the potential $\mathcal{K}^{(5th)}_{gr}(\xi)$ with no more
than fifth-order derivatives of the Killing vector, being of the form
\be
\mathcal{K}^{(5th)}_{gr}(\xi)=c_2\big({\rm d}\xi
-c_1\Box{\rm d}\xi+c_{3}\Box\Box{\rm d}\xi\big)
\, . \label{GenfifKgr}
\ee
Alternatively, it is expressed in components as
\bea
\mathcal{K}^{(5th)\mu\nu}_{gr}
&=&2c_2\Big(\nabla^{\mu}\xi^{\nu}
+c_1R^{\mu\nu}_{\rho\sigma}\nabla^{\rho}\xi^{\sigma}
+c_{3} R^{\mu\nu}_{\alpha\beta}R^{\alpha\beta}_{\rho\sigma}
\nabla^{\rho}\xi^{\sigma}\Big)\nn \\
&&-2c_2c_{3}\big(\Box R^{\mu\nu}_{\rho\sigma}\big)
\nabla^{\rho}\xi^{\sigma}
\, . \label{GennmfifK}
\eea
In order to guarantee that the integral of
$\mathcal{K}^{(5th)tr}_{gr}$ over the surface at infinity is convergent,
Eq. (\ref{GennmfifK}) shows that the constant parameters $c_1$
and $c_3$ have to be constrained through
$\big(\sqrt{-g}\mathcal{K}^{(5th)tr}_{gr}\big)\big|_{\infty}=C$, where
$C$ is a certain finite constant parameter. Because of the divergence of
$\sqrt{-g}$ at infinity for the asymptotically AdS spacetimes (this can
be seen from the metric for the AdS space, given by Eq. (\ref{SchAdS})
in the absence of the $m$ parameter), we have the necessary but not the
sufficient condition
$\mathcal{K}^{(5th)tr}_{gr}\big|_{\infty}=0$ for the fulfillment of the
convergence of the conserved charges, which is written as
\be
\Big[\big(1-2c_1\ell^2
+4c_3\ell^4\big)\nabla^{t}\xi^{r}
-c_{3}(\Box R^{tr}_{\rho\sigma})
\nabla^{\rho}\xi^{\sigma}
\Big]\Big|_{\infty}=0
\, , \label{c134con}
\ee
with the help of the falloff condition (\ref{RiemSI}).
It is easy to check that the above equation holds identically under the
following setting
\be
c_1=\frac{1}{2\ell^2} \, , \quad
c_3=0
\, . \label{c134ch}
\ee
As a consequence, under the condition that the choice of the parameters
in Eq. (\ref{c134ch}) is able to ensure that the relation
$\big(\sqrt{-g}\mathcal{K}^{(5th)tr}_{gr}\big)\big|_{\infty}=C$
holds, one can conclude that it is unnecessary to introduce the
fifth-order derivatives of the Killing vector to cancel out the divergent
terms from the usual Komar potential in the context of the Einstein gravity
with a negative cosmological constant. Obviously, such a conclusion holds
for the cases in which more than fifth-order derivatives are included in
the potentials, arising from the fact that the 2-form $\Box{\rm d}\xi$
involved in the potential $\mathcal{K}^{(3th)}(\xi)$, fulfilling
$\Box{\rm d}\xi|_{\infty}=2\ell^2{\rm d}\xi$, is proportional to the
conventional first-order derivative Komar potential at infinity so that
the linear combination ${\rm d}\xi-c_1\Box{\rm d}\xi$ disappears
inevitably at infinity under $c_1=1/(2\ell^2)$.

However, generally speaking, for a potential $\mathcal{K}^{\mu\nu}_{2N+1}$
that is the linear combination of all the $(2i+1)$-th-order (the integer
$i$ is restricted to $0\leq i\leq N$ for some given positive integer $N$)
derivatives of a Killing vector, the vanishing of $\mathcal{K}^{tr}_{2N+1}$
at infinity does not of itself guarantee that it multiplied by the factor $\sqrt{-g}$
is certainly finite at infinity. In such a case, the more than $(2N+1)$-th-order
derivatives of the Killing vector should be taken into consideration.
Anyway, our fundamental requirement is to find out the potential
$\mathcal{K}^{\mu\nu}_{2N+1}$ as simple as possible that renders
$\big(\sqrt{-g}\mathcal{K}^{tr}_{2N+1}\big)\big|_{\infty}$ finite.
For instance, if the third-order derivative potential
$\mathcal{K}^{(3th)}_{AdS}$ in Eq. (\ref{CalK3gr}) still fulfills
$\big(\sqrt{-g}\mathcal{K}^{(3th)tr}_{AdS}\big)\big|_{\infty}
\rightarrow\infty$, we can go on to consider the fifth-order derivative one
$\mathcal{K}^{(5th)}_{gr}$ in Eq. (\ref{GenfifKgr}), or the one of higher
order. As a matter of fact, in Section \ref{three}, it has been shown that
the surface integrals of the $\mathcal{K}^{(3th)}_{AdS}$ potential with
$c_1=1/(2\ell^2)$ are able to yield the physical charges of the
asymptotically AdS black holes in the context of the Einstein gravity.
To this point, it is advisable to consider only the fourth-order corrections
to the ordinary Komar current in Section \ref{one}.

\end{document}